\documentclass[12pt]{article}
\usepackage{epsfig,epsf,amssymb,latexsym}	
\usepackage{color}
\usepackage{multicol}

\begin{document}

\title{\bf Vacuum Selection on the String Landscape}
\author{Edward Tetteh-Lartey\thanks{lartey@fnal.gov} \\ 
Department of Physics, Texas $A\&M$ University, College Station, TX 77845, 
USA
}

\date{\today}

\maketitle

\small{
I examine some non-anthropic approaches to the string landscape. These approaches are based on finding the initial conditions of the universe using the wavefunction of the multiverse to select the most probable vacuum out of this landscape. All approaches tackled so far seems to have their own problems and there is no clear cut alternative to anthropic reasoning. I suggest that finding the initial conditions may be irrelevant since all possible vacua on the landscape are possible initial state conditions and eternal inflation could generate all the other vacua. We are now left to reason out why we are observing the small value of the cosmological constant (CC). I address this issue in the contest of noncritical string theory in which all values of the cosmological constant on the landscape are departures from critical equilibrium state.
}

\begin{multicols}{2}
\section{Introduction}
\paragraph{}
There is current overwhelming evidence that the universe is accelerating with a small positive cosmological constant, implying we are living in an asymptotically deSitter spacetime. There is also strong evidence that our universe must have gone through an early 
period of inflation. 

String theory, our current best theory for unifying all fundamental forces 
must be able to embed inflation and also find an explanation for the hierarchy problems facing fundamental physics. These are:
\begin{itemize} 
\item The difference in scale between the weak scale and the Planck scale 
\item The small value of the cosmological constant 
\item The strong CP problem.
\end{itemize}

I address the inflationary issue in this review and leave the hierarchy problem in my next paper.

To explain our universe we need a scalar field rolling down its potential to a dS vacuum state, in such a way that the potential is flat enough in the early stages to achieve at least 60 e-folds of inflation as required by observation. We also need inflation to end at some point and only begin later when the vacuum energy dominants the matter and radiation energy densities.

Obtaining an inflationary potential with a dS vacuum state has been a difficult issue in string theory for a while. The reasons being:
\begin{itemize} 
\item The no-go theorem which guarantees that no deSitter solutions can be obtained in string/M theory by using only the lowest order terms in the 10d or 11d supergravity action$~\cite{no-go}$. 
\item String compactifications to our 4d universe comes with a large number of moduli fields which control the size and shape of the compactification manifold as well as the string coupling. Inflation is only possible if these fields are either stable or else have relatively flat potentials which do not cause fast, non-inflationary rolling in field space. To generate inflation one should learn how to stabilize these moduli which is not an easy thing to do. Thus for a long time it had been seemed very difficult to obtain inflation from String/M Theory.  
\end{itemize}

A solution to these problems was found in$~\cite{kklt}$ (KKLT construction). The authors showed that by using quantum corrections and extended objects in string theory, the no-go theorem could be circumvented to achieve dS solutions in critical string theory. DeSitter solutions were also shown to be achievable in noncritical string theory in$~\cite{noncrit}$. A dS vacua was obtained in the KKLT scenario by three steps: 1) Fluxes were turned on a type IIB orientifold compactified on a 3-fold Calabi-Yau manifold to stabilize all but the volume modulus and breaking supersymmetry. 
2) Non-perturbative effects were then introduced stabilizing the volume modulus, restoring supersymmetry and producing an AdS vacuum. 3) $\overline{D3}$ were then introduced in this warped geometry, uplifting the AdS vacuum to a dS vacuum and breaking supersymmetry.   

Having achieved a dS vacuum state, the next goal is to achieve inflation with at least 60 e-folds as I stated earlier. The potential obtained in the KKLT scenario was however not naturally flat to achieve the desired number of e-folds of inflation. 

Following flux compactification and moduli stabilization scenario in the KKLT construction, the authors of$~\cite{kklmmt}$ (KKLMMT) obtained inflation by introducing a $D3-\overline{D3}$-brane pair, whose vacuum energy drives the inflaton breaking supersymmetry explicitly.
Since then other ways of obtaining inflation have been found including introduction of a $D3-D7-$brane pair with fluxes, resulting in the appearance of a D-term and leading to spontaneous breaking of supersymmetry$~\cite{dterm}$. 
In KKLT scenario, choices of flux and Calabi-Yau manifold leads to discrete values of the vacuum energy, demonstrating a vast landscape of plausible solutions of string vacua called the ``String landscape''. A first example of the vast landscape of string theory was shown in$~\cite{bousso}$. The authors demonstrated that by using multiple four form gauge fluxes arising from M theory compactifications on manifolds with non-trivial three-cycles and following Dirac quantization rules, one could generate of order 100 discrete metastable vacaus of cosmological constants. This makes it possible for the observed value of the cosmological constant $\Lambda\approx 10^{-120}M_{PI}^{4}$ to be plausible as
one of these solutions without any need of fine-tuning.
Work by$\cite{douglas}$ on vacuum statistics showed that there are about $10^{500}$ string vacua solutions populating the landscape. 
The demerit of this vast landscape is that it leads to a vacuum selection problem of why we live in this particular vacuum out of the many possibilities. 
The anthropic principle\footnote{It states that we are observing this value of the cosmological constant being that it is what is well tuned for life. Any value much higher will not be suitable for life and we will not be here to observe it.} 
has been suggested as a viable explanation for why we are observing the small value of the cosmological constant. But this has been an uncomfortable zone for a lot of physicist since it is not falsifiable and also difficult to make predictions from. Finding physical quantities that string theory might be able to predict becomes a very hard or impossible task, at least until probability distributions peaked around universes like ours can be sensibly derived.

There have been a number of attempts to find an alternative to the anthropic principle by constructing the wavefunction of a multiverse. This wavefunction is dependent on variables describing the landscape. The probability distribution from this wavefunction is then used to make predictions for the values of these parameters and hence the most likely vacua to expect. 
By this way we can understand the initial conditions of the universe in addition to understanding the dynamic evolution of the universe from string theory. The string landscape also provides driving influence in pursing this direction.
 
The Hartle-Hawking wave function of the universe favors a universe with cosmological constant $\Lambda 
= 0$. The wavefunction is unbounded below and hence not normalizable. Also $\Lambda=0$ implies 
inflation cannot occur which is a clear contradiction to observation.
Work in$~\cite{tye1,tye2}$ showed that by making corrections from decoherence effects the 
wavefunction is modified and bounded below with the probability distribution peaking at vacuas with 
intermediary energies away from zero. 
In$~\cite{laura}$ it was pointed out that the correction just renormalizes the wavefunction and that 
the problem of normalisability still remains. 
An alternative to the Hartle-Hawking wavefunction is the Linde$~\cite{linde}$ and 
Vilenkin$~\cite{vilenkin}$ wave function but this is also not bounded above and seems to be in conflict 
with WMAP results on the value of the Hubble parameter.

In$~\cite{laura2}$ a vacuum selection idea was proposed based on 
dynamic selection of the most probable vacua solution on the landscape superspace.
The problem with all these approaches is that even if the most probable dS vacuum can be selected from the landscape, a dS vacuum is a problem for string theory and a correct mathematical framework for addressing this issue is required.

The question is what observables do we
define in a dS spacetime in string theory. A dS spacetime has a cosmological horizon, and observers can
only access a fraction of information in this spacetime and the S-Matrix is not well defined. Also due to
information loss across the horizon, unitarity is violated and it is impossible to define pure
asymptotic states, which are essential for the proper definition of a scattering S-Matrix.
Note that in conventional quantum theory the S-Matrix connects asymptotic well separated 'in' states to asymptotic well separated 'out'
states by:

\begin{equation}
|out> = S|in>
\end{equation}

However if we encounter mixed states due to the presence of the horizon, then the concepts of "in" and
"out states" should be replaced by those of "in" and "out density matrices", given that pure states
evolve
to mixed ones. 
In such a situation the "in" and "out'' density matrices can be linked by a $\$$ Matrix instead of an
S-Matrix

\begin{equation}
\rho_{out} = \$\rho_{in}
\end{equation}

Due to unobserved degrees of freedom, the operator $\$$ loses its factorisability: $ \$ \neq SS^{\dag}$\footnote{
The operator $\$$ factorizes into a product $SS^{\dag}$ only in pure state quantum mechanics without
unobserved degrees of freedom, in which $\rho = |\psi><\psi|$. In general, however, once there are
unobserved degrees of freedom, thereby opening up the system, as is the case of horizons (local or
global), the factorization property of $\$$ lost: $ \$ \neq SS^{\dag}$
}
and on-shell scattering amplitudes cannot be defined.
Even if we invoke models for relaxing the vacuum energy, leading asymptotically to a vanishing vacuum energy density, i.e a Minkowski spacetime which will be consistent with an S-Matrix, as was shown in KKLT scenario, where the dS spacetime was a metastable state which should decay to a Minkowski vacuum at a time less that the poincare recurrence, there is still an open issue of embedding such models in (perturbative) string theory. 

All suggested methods based on finding the initial conditions of the universe seem to have their own problems.
In this review I take a different approach. I suggest that any vacuum state in the landscape could be a possible starting point of the universe. Eternal inflation will then generate all other possible vacua on the landscape. 
I then embed this vacau in a noncritical string framework by considering each as a  metastable non-conformal, non-equilibruim state indicating a departure from
criticality with a central charge deficit which then flows down to a conformal point in theory space.

In section 2, I look at some approaches on finding the most probable 
vacau on the landscape using the wavefunction of the multiverse and address some flaws that come with these approaches. I discuss the framework of non-critical string theory and its application to cosmology in section 3. In section 4, I discuss my alternative approach to using the initial conditions of the multiverse by considering rather an eternally inflating universe driving towards a conformal point in theory space. I conclude in section 5.
 
\section{Vacuum Selection Mechanism}
\paragraph{}
The Hartle-Hawking wavefunction is expressed as a path integral over compact Euclidean geometries bounded by a given 3-geometry g:

\begin{equation}
\Psi_{HH(g,\varphi)} = \int^{(g,\varphi)}e^{-S_{E}}  \label{eqwfn1}
\end{equation}

where $S_{E}$ is the Euclidean action which gives a measure of the entropy and $\Psi$ satisfies the Wheeler-DeWitt equation:

\begin{equation}
\hat{H}\Psi_{HH(g,\varphi)} = 0  \label{dwhe},
\end{equation}

analogous to the Schrodinger equation in quantum mechanics.

On assuming that the dominate contribution to the path integral is given by the stationary points of the action we may approximate this wavefunction $\Psi_{HH(g,\varphi)}$ simply as:

\begin{equation}
\Psi_{HH} \sim e^{-S_{E}} = \exp\left(\frac{3\pi}{2G_{N}\Lambda}\right) \label{eqwfn2} 
\end{equation}

where $\Lambda$ is the cosmological constant and $G_{N}$ is Newton's gravitational constant.
This gives the probability for a universe to spontaneously nucleate out of 'nothing'\footnote{By nothing I mean no classical space and time} to a closed deSitter spacetime with cosmological constant $\Lambda$, and given by:

\begin{equation}
P_{HH} = |\Psi_{HH}|^{2} \simeq \exp\left(\frac{3}{8\Lambda} \right) \label{amp1}
\end{equation}

We can obviously see that the wavefunction indicates a preference for low cosmological constant with a peak at $\Lambda = 0$. The wavefunction is unbounded below and thus non-normalizable.
Prediction from this wavefunction is that the universe will nucleate to a non-inflationary universe. This clearly contradicts our observation of an inflationary universe.  
However by including matter fluctuations and metric perturbations, which tend to decohere the wave function as proposed by Tye et al$~\cite{tye1, tye2}$,
the tunneling amplitude is suppressed. Thus in the string landscape scenario where different vacua are generated dependent on the choices of nontrivial fluxes and branes-antibranes this suppression will be dependent on these flux and brane configurations and a prediction of a preferred compactification can hence be made.  The wavefunction then modifies from $(~\ref{eqwfn2})$ to:

\begin{equation}
\Psi_{soup} \sim e^{S_{soup}}, \     S_{soup} = -S_{E} - {\it D} \label{eqwfn3}
\end{equation}

where {\it D} is the decoherence correction term. Tye et al obtained a tunnel probability with this correction term given by

\begin{equation}
P_{soup} \sim  e^{F_{soup}} = \exp\left(\frac{3}{8\Lambda} - \frac{27\nu}{32\Lambda^{2}}\right) \label{ampsoup}  
\end{equation}

where $\nu = M^{4}/(72\pi^{6})$ \footnote{$\nu$ is a constant which measures the number of perturbative modes and depends on the UV cutoff for the gravitational interaction in a quantum theory of gravity. In string theory the UV cutoff is chosen to be the string scale $M_{s}$.}, M being the UV cutoff scale which for string theory is chosen to be the string scale $M_{s}$. Thus the actual value of $\nu$
cannot be calculated until we know the string scale. The authors then hoped that this modified function will suppress the tunneling rate and select a $\Lambda$, other than the one selected by the original HH function, with $\Lambda \rightarrow 0$ as the most probable value and thus bound the Euclidean action from below. This proposal has however been refuted by$~\cite{laura}$, stating that the correction term just renormalizes the cosmological constant and thus the problem of non-normalizable wavefunction remains unsolved.  

The Linde$~\cite{linde}$ and 
Vilenkin$~\cite{vilenkin}$ wave function proposal gives a probability distribution:

\begin{equation}
P_{L,V} \sim e^{-\frac{3}{8\Lambda}}
\end{equation}

This clearly favors an inflationary universe with a large $\Lambda$. The 
problem is that it predicts a Hubble parameter $H \sim \sqrt{M_{s}^{4}/M_{p}^{2}}$.
If we should take the string scale to be an order of magnitude below the Planck scale\footnote{This is the case for string compactification where there are no anomalously large extra dimensions} then $H\sim 10^{16}GeV$. This falls in conflict with WMAP results that $H\leq10^{14}GeV$, unless the cutoff scale is two orders of magnitude below the Planck scale.  

So it appears that finding initial conditions of the universe as a way of selecting the most probable vacua runs us into another set of problems. But 
inflation can help in this search for a non-anthropic explanation of vacuum selection.
The string landscape scenario makes eternal inflation a high possibility since it can provide a mechanism to populate the landscape with all possible values of CC. If inflation is eternal then the concept of initial conditions are irrelevant. In an eternally inflating universe anything that can happen will happen and will happen an infinite number of times. The question now is how to define probabilities in eternally inflating spacetimes. Defining these probabilities leads to ambiguities$~\cite{guth}$.

In$~\cite{laura2}$ the idea was to consider an array of vacua solutions as a lattice allowing the 
wavefunction of the universe to propagate on this landscape background. This system is then studied as a quantum N-body system. A probability distribution can then be calculated from the solutions
 of the Wheeler-DeWitt equation (WDW) for the wavefunction, and the most probable universe predicted from the peak. The vacua are parametrized by a collective coordinate $\phi$ with potential $V(\phi)$, where $\phi$ is a moduli field collectively characterizing all the internal degrees of freedom in each vacua, but taking distinct values in each vacua. Thus the vacua energies will be labeled as $V(\phi_{i})=\lambda_{i}$, i = 1, 2, 3,.....N, where N = number of vacua in landscape. A minisuperspace is defined on a superspace restricted to this landscape ``lattice'' and parametrized by the collective coordinate $\phi$ and, to homogeneous and flat 3-geometries with scale factor a(t). The metric is given by:

\begin{equation}
ds^{2} = -{\it N}dt^{2} + a^{2}(t)dx^{2} \label{metric}
\end{equation}

where ${\it N}$ is the lapse function set to one. The wavefunction $\Psi(a,\phi)$ then propagates on this minisuperspace spanned by $(a,\phi)$.
The Wheeler-De Witt equation is then given by:

\begin{equation}
\hat{H}\Psi(a,\phi) = 0,
\end{equation} 

\begin{equation}
\hat{H} = \frac{1}{2e^{3\alpha}}\left[ \frac{4\pi}{3M_{p}^{2}}\frac{\partial^{2}}{\partial\alpha^{2}} - \frac{\partial^{2}}{\partial\phi^{2}} + \exp^{6\alpha}V(\phi) \right],  \label{wdw}
\end{equation}

where a is replaced by $a=e^{\alpha}$.

Rescaling $\phi$ to $x = e^{3\alpha}\phi$ and decomposing $\Psi(a,\phi)$ into modes 

\begin{equation}
\Psi(\alpha, x) = \sum_{j}\psi_{j}(x)F_{j}(\alpha) \label{wdw2}
\end{equation}

then substituting into $(~\ref{wdw})$ gives:

\begin{equation}
\frac{3M_{p}^{2}}{4\pi}\left[ \frac{\partial^{2}}{\partial x^{2}} - V(X) \\ 
\right]\psi_{j}(x) = e^{6\alpha}\epsilon_{j}\psi_{j}(x)  \label{wdw3}
\end{equation}

resulting in
\begin{equation}
-\frac{\partial^{2}}{\partial\alpha^{2}}\psi_{j}(x)F_{j}(\alpha) = -\epsilon^{\prime}_{j}\psi_{j}F_{j} \label{wdw4}
\end{equation}

where $\epsilon^{\prime}_{j} =  e^{6\alpha}\epsilon_{j}$ with $\frac{3M_{p}^{2}}{4\pi}$ absorbed into $\epsilon_{j}$.

By solving the Schrodinger equations above, eigenvalues $\epsilon^{\prime}_{j}$ are obtained for the field $\psi_{j}(x)$ propagating on the superlattice V(x) with N lattice sites $x_{i}$, and vacua energies $\lambda_{i}$.

To obtain the equations of motion for $\alpha$, we vary the action $S(\alpha, \phi)$ in $(~\ref{wdw})$ with respect to $\alpha$

\begin{equation}
\ddot{\alpha} + \frac{3}{2}\left[ \dot{\alpha}^{2} + (\dot{x}^{2} - V(x))e^{-6\alpha} \right] = 0 \label{wdw5}.
\end{equation}

A claim by the author$~\cite{laura2}$ is that, it is the Friedman equation for the expansion of the universe born out of the wavefunction $\Psi(\alpha, \phi)$ solutions to the WDW equation.

The author$~\cite{laura2}$ then goes on to calculate the most probable universe for both the SUSY and non-SUSY sectors of the landscape. The most probable universe in the non-SUSY sector gave $\Lambda = 0$. This appears inconsistent with inflationary universe.

I take a detour at this point by suggesting a difference approach where finding the initial conditions is irrelevant. The approach is based on noncritical string theory. I conjecture that the non-SUSY sectors with non-zero vacuum energies indicate non-equilibrium points on the string world sheet. Thus there is an edge at each lattice site to restore conformal invariance. The site with the most high energy vacuum energy is the most unstable. 

I discuss noncritical string framework in the next section and its application to the landscape in section 4.

\section{Noncritical string and Liouville Theory}
\paragraph{}
Critical string theory in which target spacetime is 10 or 11 for Superstring/M Theory and 26 for bosonic string theory is conformally invariant only in flat spacetime with the dimensions stated for consistency with Lorentz symmetry. Thus this describes only fixed background situations. However a theory which is suppose to describe spacetime and how it was created must be background independent. This is obviously a high point for the closest competitor, loop quantum gravity, 
which is a background independent theory describing a dynamic spacetime. A way around this is to get rid of conformal invariance which fixes the background and 
use instead non-critical (Liouville) strings, which is a mathematically consistent way to deal with $\sigma$-models away from their conformal (fixed) points on the space of string theories (see fig 1). 

\begin{center}
\includegraphics[width=4.0in]{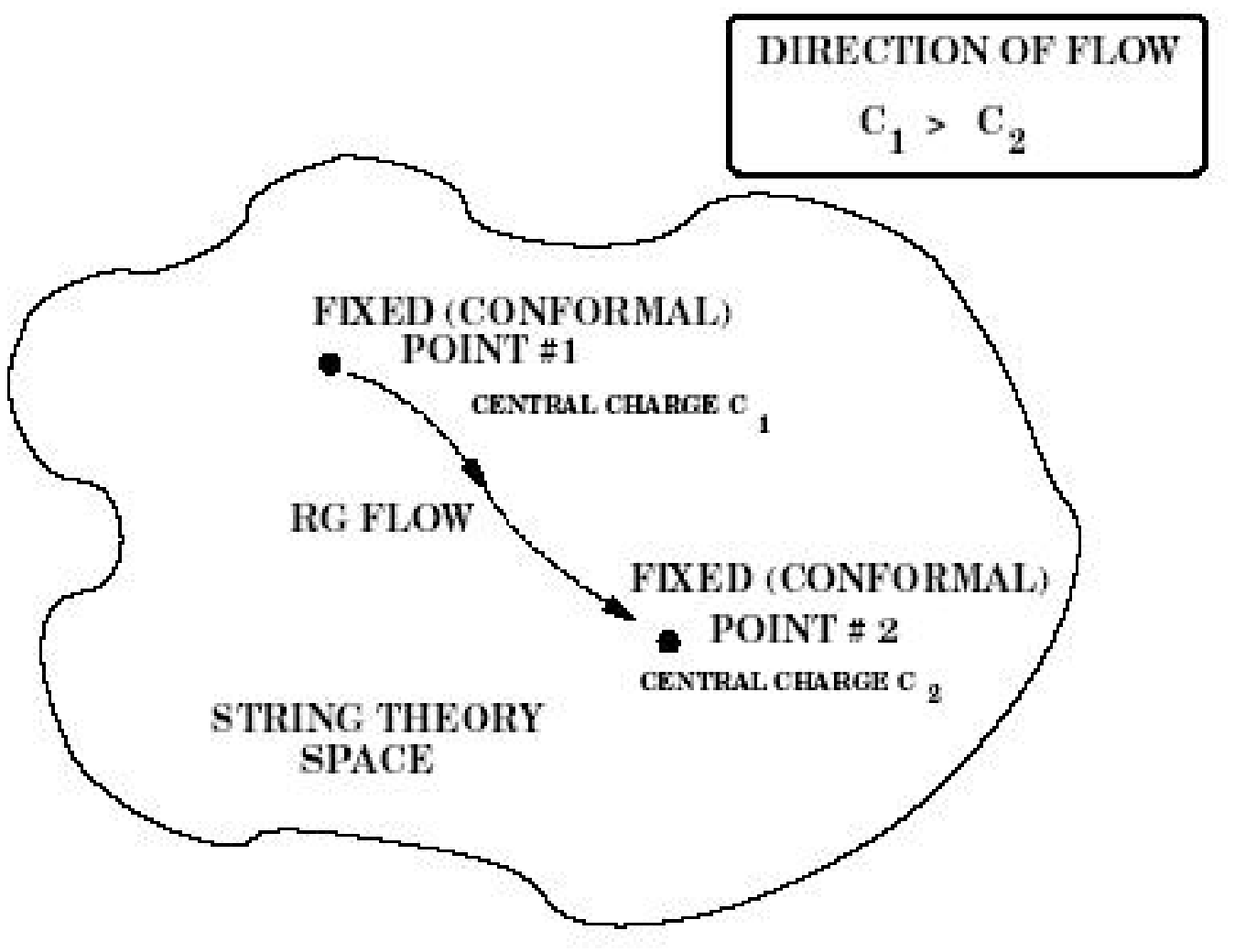} 
{\it Fig. 1. A schematic view of string theory space. This is an 
infinite-dimensional manifold endowed with a (Zamolodchikov) metric. 
The dots denote conformal string backgrounds. 
A non-conformal string flows (in a two dimensional renormalization-group sense) from one fixed point to another, either of which could be a hypersurface in theory space. The direction of the flow is irreversible, and is directed towards the fixed point with a lesser value of the central charge, for unitary theories, or, for general theories, towards minimization of the degrees of freedom of 
the system$~\cite{mav}$.}
\end{center} 

With noncritical strings theory we may describe arbitrary number of backgrounds with no limit on the dimensionality of spacetime, which we call a theory space. This is considered as the space of all possible background target spacetime string configurations, an infinite dimensional manifold. The manifold is endowed with a Zamolodchikov metric given by the appropriate two point function of the vertex operators describing the non-conformal deformations: $G_{ij} =|z\overline{z}|^{2}<V_{i}(z,\overline{z})V_{j}(0,0)>$, where the vertex operators are deforming the $\sigma$-model as:

\begin{equation}
S_{\sigma} = S^{*} + \int_{\sum}g^{i}V_{i} \label{sigma} 
\end{equation}

where $S^{*}$ is a fixed point $\sigma$-model background action (a conformal equilibrium point) and $g^{i}$ are the non-conformal backgrounds/couplings corresponding to the vertex operators $V_{i}$. For example the $\sigma$-model action in a background with graviton G, antisymmetric tensor B and dilaton $\Phi$ reads $~\cite{mgreen}$:

\begin{eqnarray*}
S_{\sigma} & = & \frac{1}{4\pi\alpha^{\prime}}\int_{\sum}d^{2}
\xi[\sqrt{-\gamma}G_{\mu\nu}\partial_{\alpha}X^{\mu}\partial^{\alpha}X^{\nu} 
\end{eqnarray*}
\begin{eqnarray*}
&  &+ i\epsilon^{\alpha\beta}B_{\mu\nu}\partial_{\alpha}
X^{\mu}\partial_{\beta}X^{nu}
\end{eqnarray*}
\begin{eqnarray}
&  &+ \alpha^{\prime}\sqrt{-\gamma}R\Phi] \label{sigma2}
\end{eqnarray}

where $\sum$ denotes the 2D world sheet, $\gamma$ the world sheet metric, $\alpha$, $\beta$ are world sheet indices, $\mu, \nu$ are target spacetime indices, and R is the 2D scalar curvature of the world sheet.

Note that target spacetime fields, $G_{\mu\nu}$, $B_{\mu\nu}$, and $\Phi$ appear as couplings to the 2D observer which I denote collectively as $g^{i}=(G_{\mu\nu}, B_{\mu\nu},\Phi)$.    

The dilaton is define as:

\begin{equation}
\Phi = const - \frac{1}{2}Q(t)t \label{sigma3}
\end{equation}

Q is called the central charge deficit. It decreases in value with time as it flows down to the conformal constant dilaton zero point. 

\begin{eqnarray*}
Q^{2} & = & 0, \ we \ have \ criticality \\
Q^{2} & > & 0, \ we \ have \ supercriticality \\
Q^{2} & < & 0, \ we \ have \ subcriticality \label{qe}
\end{eqnarray*}  

$S^{*}$ is the flat spacetime action where the target spacetime fields (or world sheet couplings take the values;

\begin{equation}
G_{\mu\nu} = \eta, \ B_{\mu\nu}=0, \  and  \ \Phi = constant \ or \ zero, \label{sigma4}
\end{equation}

called the linear dilaton background and is a conformal fixed point.

The second term in $(~\ref{sigma})$ is the deformation term indicating a departure from criticality. 
The graviton world sheet $\beta$ function in the absence of any other background is given by:
\begin{equation}
\beta_{\mu\nu} = \alpha^{\prime}R_{\mu\nu} \label{grav}
\end{equation} 

In this situation conformal invariance in the action $eq(~\ref{sigma}$) is only possible if $\beta_{\mu\nu}=0$, i.e it allows only vacuum solutions to the Einstein equations. $\beta$ functions of all other fields are also required to be zero for conformal invariance. This non-conformality implies $\beta^{i} \equiv dg^{i}/d\ln\mu \neq 0$, where $\mu$ is the world sheet renormalization scale. Thus conformal invariance implies restrictions on the background and couplings $g^{i}$.

Closed to the fixed conformal points, the $\beta^{i}$ function can be expanded perturbatively in terms of the coupling $g^{i}$: $\beta^{i} = y_{i}g^{i} + c^{i}_{jk}g^{i}g^{k}$ + ...., where no sum is implied in the first term, $y_{i}$ is the anomalous dimension, and $c^{i}_{jk}$ are the operator product expansion (OPE) coefficients.

These off-shell $\beta$ functions can be derived as $g^{i}$-space gradients of a flow function C[g,t], which is a renormalization-group invariant function coinciding with the running central charge of the theory between fixed points. This is the Zamolodchikoz c-theorem which establishes the relation:

\begin{equation}
\frac{\partial C[g,t]}{\partial t} = \frac{-1}{12}\beta^{i}G_{ij}\beta^{j}, 
\ \beta^{i} = G_{ij}\frac{\partial C[g,t]}{\partial g^{j}} \label{cth}
\end{equation} 

The function C[g,t] acts like a thermodynamic H function, decreasing monotonically along the direction of flow and is a measure of the central charge deficit. At fixed conformal points $\beta^{i} = G^{ij}\frac{\partial C[g,t]}{\partial g^{j}} = 0$.

At non-conformal points in the gradient flow where $\beta^{i}\neq0$, the theory is in need of dressing by a Liouville field $\phi$ in order to restore conformal symmetry$~\cite{mav}$. Note that in the non-conformal background the Liouville mode does not decouple from the action and the Liouville action is written in terms of the dresses couplings $g^{i}(\phi)$ as:

\begin{eqnarray*}
S_{L} & = & S^{*} + \frac{1}{8\pi}\int_{\sum}d^{2}\xi\sqrt{\hat{\gamma}}
[\pm(\partial\phi)^{2} - QR\phi] 
\end{eqnarray*}
\begin{eqnarray}
& &+ \int_{\sum}d^{2}\xi\sqrt{\hat{\gamma}}g^{i}(\phi)V_{i} \label{laction}
\end{eqnarray} 

where $\hat{\gamma}$ is a fiducial world-sheet metric, and the plus(minus) sign in front of the kinetic terms of the Liouville mode indicate subcritical(supercritical) strings. The dressed couplings $g_{l}^{i}(\phi)$ are given by:

\begin{equation}
g_{l}^{i}(\phi) = g^{i}e^{\alpha_{i}\phi} + \frac{\pi\phi}{Q\pm2\alpha_{i}}c^{i}_{jk}g^{j}g^{k}e^{\alpha_{i}\phi} \label{ldress}
\end{equation}

Liouville dressing leads to an increase in the target spacetime dimensionality from D to D+1. The equation relating the Liouville dressed couplings $g_{l}^{i}$, the $\beta$ functions and the central charge deficit Q are given by:

\begin{equation}
\ddot{g_{l}}^{i} + Q(t)\dot{g_{l}}^{i} = \mp \beta^{i} \label{leq2}
\end{equation} 

where the dot implies derivative with respect to the Liouville world-sheet zero mode. 
In a general setup of strings propagating in background fields $G_{\mu\nu}$, $B_{\mu\nu}$, and $\Phi$, the $\beta$ functions of the fields in D-dimensional spacetime are given:

\begin{eqnarray}
\beta^{G}_{\mu\nu} & = &\alpha^{\prime}\left( R_{\mu\nu} 
  +2\nabla_{\mu}\partial_{\nu}\Phi - \frac{1}{4}H_{\mu\rho\sigma}H^{\rho\sigma}_{\nu}  \right) \\
\beta^{B}_{\mu\nu} & = & \alpha^{\prime}\left( -\frac{1}{2}\nabla_{\rho}H^{\rho}_{\mu\nu} + H^{\rho}_{\mu\nu}\partial_{\rho}\Phi \right) \\
\beta^{\Phi} & = & \frac{1}{6}\left(D 
- \frac{3}{2}\alpha^{\prime}\omega - 26 \right) \label{beta}
\end{eqnarray}
\begin{eqnarray*}
\omega & = & [R - \frac{1}{12}H^{2} - 4(\nabla\Phi)^{2} 
 +4\Box\Phi]
\end{eqnarray*}

where $\alpha^{\prime}$ is the Regge slope and $H_{\mu\nu\rho} = \partial_{[\mu} B_{\nu\rho]}$ is the field strength of the B field. The indices $\mu,\nu = 0....D-1$.

We may however identify the Liouville zero mode with time $\phi = -t$, thereby reducing the spacetime dimensionality from D+1 to the critical dimension D. In the supercritical case (positive central charge deficit) an inflationary 
universe, the Liouville conformal equation would be written as:

\begin{equation}
\ddot{g_{l}}^{i} + Q(t)\dot{g_{l}}^{i} = \mp \beta^{i}(g) = -G^{ij}\partial C[g] /\partial g^{j} \label{leq3}
\end{equation}

where the dot implies derivative with respect to target time since we have identified it with the Liouville zero mode $\phi$, and $G^{ij}$ is the inverse Zamolodchikov metric.

Here $Q^{2}(t)$ is related to the vacuum energy and relaxes slowly to zero with time. It is constant in the inflationary era and related to the Hubble parameter H by:

\begin{equation}
Q^{2} = 9H^{2} > 0 \label{vener}
\end{equation}  

\begin{equation}
\Lambda \propto Q^{2}(t) \label{lam}
\end{equation}

Thus any vacuum energy generated on the string landscape can be related to $Q^{2}$ which then relaxes to zero conformal point. 

\section{Discussions}
\paragraph{}
I interpret the landscape as a result of instability in a subsection of the superspace. This instability is in two parts:
\begin{itemize}
\item Destabilization of the superspace:  
This created the dynamic structures, branes, fluxes, strings etc. 
\item Destabilization of the conformal world sheet of the dynamic structures: 
\end{itemize}

The destabilization of the Superspace may be a result of quantum fluctuations or spontaneous breaking of a symmetry yet to be discovered. The result of this is the creation of substructures from the manifold such as branes, fluxes, strings etc. The dynamics of this substructures and wrappings on the manifold created a potential leading to the stabilization of the moduli fields and breaking of conformal symmetry on the world sheets of these structures and triggering inflation on the branes.
A single configuration of branes, fluxes with a positive vacuum energy from the landscape and right conditions for inflation could have been the initial state of the universe. Once inflation started it became eternal, creating the landscape of cosmological constants. Due to this eternal inflation, finding the initial conditions is not necessary since any one vacua could create the landscape of pocket universes. A universe with the largest energy density would be the most unstable with the largest drive to restore conformal symmetry.
One could temporarily restore conformality by Louiville dressing. This leads to an increase in dimensionality of the spacetime. But dimensionality of the superspace should be fixed for all time. This can be achieved by identifying the Louiville mode with target time. In summary we achieve a temporal restoration of conformal invariance without necessarily increasing the spacetime dimensions. 

There may be a tendency for permanent restoration of conformal symmetry. This forces all these pocket universes to merge into one and move towards a conformal zero point on the string theory space. This move is discretized i.e, takes the values on the landscape in decreasing order. We may be witnessing the stage where the value it has taken is suitable for life and it is what we have measured as the cosmological constant.

This destabilization may only have occurred in a subsection of the giant Superspace. We may be witnessing a permanent restoration of the conformal symmetry which was broken. The destabilization of the superspace may be restored by annihilation of the branes, fluxes and all the other structures back into the superspace.

\section{Conclusions}
\paragraph{}
I have examined some non-anthropic approaches to the string landscape. These approaches are based on finding the initial conditions of the universe using the wavefunction of the multiverse to select the most probable vacuum out of this landscape. All approaches tackled so far seems to have their own problems and there is no clear cut alternative to anthropic reasoning.

I suggest that finding the initial conditions may be irrelevant since all possible vacua on the landscape are possible initial state conditions and eternal inflation could generate all the other vacua. We are now left to reason out why we are observing the small value of $\Lambda$. I address this issue in the context of noncritical string theory in which all values of the CC on the landscape are departures from critical equilibrium state. Each of these vacua has a central charge deficit which flows down to a conformal zero point on the string theory space. The rate of flow may be proportional to the degree of departure such that at some point they all merge into one to flow down the path. We are probably witnessing the point where the value is suitable for life and it is what we measure as the CC.
I also mentioned that branes, fluxes and strings might have been created out of this giant Superspace as a result of spontaneous breaking of a symmetry yet to be discovered or a quantum fluctuation on this Superspace.


\end{multicols}

\end{document}